\documentclass{aa}
\usepackage{graphicx}
\usepackage{journals}
\usepackage{natbib}
\def\HI{{H\,{\small I}}}
\def\HIt{{H\,{\scriptsize I}}}
\def\HIb{{H\,{\Large I}}}

\begin{document}

\title{The origin of \HIb-deficiency in galaxies on the outskirts of the
Virgo cluster}
\subtitle{I. How far can galaxies bounce out of clusters?}
\titlerunning{\HIt-deficiency on outskirts of Virgo. I}

\author{
  G. A. Mamon \inst{1,2}
          \and
T. Sanchis \inst{3}
\and
          E. Salvador-Sol\'e\inst{3,4}
          \and
          J. M. Solanes\inst{3,4}
	  }

\offprints{G. Mamon, 
\email{gam@iap.fr}}

\institute{Institut d'Astrophysique de Paris (CNRS UMR 7095), 98 bis Bd Arago,
F-75014 Paris, France\\
\email{gam@iap.fr}
\and
GEPI (CNRS UMR 8111), Observatoire de Paris, F--92195 Meudon Cedex, France
\and
Departament d'Astronomia i Meteorologia, Universitat de Barcelona, Mart\'{\i} i
Franqu\`es 1, 08028 Barcelona, Spain\\
\email{tsanchis@am.ub.es; eduard@am.ub.es; jsolanes@am.ub.es}
\and 
CER on Astrophysics, Particle Physics, and Cosmology, Universitat de Barcelona,, Mart\'{\i} i
Franqu\`es 1, 08028 Barcelona, Spain
}

\date{Received 1 August 2003 / Accepted 23 October 2003}

\abstract{Spiral galaxies that are deficient in neutral Hydrogen are observed
on the outskirts of the Virgo cluster. If their orbits have crossed the inner
parts of the cluster, their interstellar gas may have been lost through ram
pressure stripping by the hot X-ray emitting gas of the cluster. We estimate
the maximum radius out to which galaxies can bounce out of a virialized system
using analytical arguments and cosmological $N$-body simulations. In
particular, we derive an expression for the turnaround radius in a flat
cosmology with a cosmological constant that is simpler than previously derived
expressions. We find that the maximum radius reached by infalling galaxies as
they bounce out of their cluster is roughly between 1 and 2.5 virial radii.
Comparing to the virial radius of the Virgo cluster, which we estimate from
X-ray  observations, these \HIt-deficient galaxies appear to lie significantly
further away from the  cluster center. Therefore, if their distances to the
cluster core are correct, the  \HIt-deficient spiral galaxies found outside of
the Virgo cluster cannot have lost their gas by ram pressure from the hot
intracluster gas. 

 }

\maketitle

\section{Introduction}    
\label{intro}
Radio observations at 21cm have revealed that spiral galaxies within clusters
are deficient in neutral Hydrogen \citep*[e.g.][]{CBG80}, and their
\HI-deficiency, normalized to their optical diameter and morphological type, is
largest for the spirals near the cluster center
\citep{HG86,CvGBK90,Solanes+01}. \cite{CBG80} suggested that the \HI-deficiency
of cluster spirals was caused by the ram pressure stripping of their
interstellar Hydrogen by the hot intracluster gas that emits in X-rays. 
Galaxies falling face-on into a cluster experience a ram pressure that scales
as $\rho_\mathrm{cl}\,v^2$ \citep{GG72}, where $\rho_\mathrm{cl}$ is the
cluster gas density and $v$ is the relative velocity of the spiral galaxy in
its cluster. Therefore, ram pressure stripping requires the large infall
velocities present in rich clusters.

\cite{Solanes+02} recently discovered deficient spirals in the periphery of the
Virgo cluster, with several ones typically over 5 Mpc in front or behind the
cluster core. In an ensuing study, \cite{Sanchis+02} could not discard the
possibility that some of these galaxies could have passed through the cluster
core and in the process had their interstellar gas swept out by the ram
pressure caused by the intracluster hot diffuse gas.

The idea of galaxies beyond the virial radius having passed through the main
body of a cluster in the past has been addressed by \cite*{BNM00} in the
context of the discovery of reduced star formation rates on the outskirts of
clusters in comparison with the field
\citep{Balogh+97}.
Using cosmological simulations,
\citeauthor{BNM00} analyzed 
6 clusters within a sphere of 2
times their final virial radius and found that $54\pm20\%$ of the particles
between $r_{200}$ and $2\,r_{200}$ (where $r_{200}$ is the radius where the
mean density of the cluster is 200 times the critical density) have actually
been inside the virial radius of the main cluster progenitor at some earlier
time. 
Unfortunately, \citeauthor{BNM00} do not provide any precision on the maximum
distances that such particles that were 
once within a cluster progenitor can move out
to. Furthermore, one needs to check if
$2\,r_{200}$ represents a sufficient distance for particles bouncing out of
virialised structures to explain the \HI-deficient
galaxies on the outskirts of the Virgo cluster.

In this paper, we ask whether the \HI-deficient galaxies on the outskirts of
the Virgo cluster  have previously passed through the core of the cluster,
using both analytical arguments and the output of cosmological $N$-body
simulations. In Sect.~\ref{nbody}, we describe the $N$-body simulations
analyzed in this paper. Next, in Sect.~\ref{phase}, we study the structure in
radial phase space of dark matter halos of the simulations. In
Sect.~\ref{maxreb}, we compute the maximum rebound radius, both analytically,
making use of the turnaround radius of cosmological structures, which we
compute in an appendix, and by studying the structure of our simulated halos in
radial phase space, as well as analyzing the orbital evolution of particles in
the cosmological simulations of \citet{FM01}. In Sect.~\ref{virVirgo}, we
estimate the virial radius and other virial parameters of the Virgo cluster to
permit the estimation of the Virgo rebound radius in physical units. We discuss
our results in Sect.~\ref{disc}.

In a companion paper \citep{SMSS03}, we discuss in more detail the origin of
the \HI-deficiency in galaxies on the outskirts of the Virgo cluster, by
analyzing 2D slices of the 4D phase space (right ascension, declination,
distance and radial velocity) and comparing them with the cosmological $N$-body
simulations used here.

\section{$N$-body simulations}
\label{nbody}

The $N$-body simulations used here to find $N$-body replica of the Virgo
cluster were carried out by \citeauthor{Ninin99} (\citeyear{Ninin99}, 
see \citealp{Hatton+03}) in the
context of 
the {\tt GALICS} (\citeauthor{Hatton+03}) hybrid 
$N$-body/semi-analytic model of hierarchical galaxy formation.  
% The outlines of
% this model can be found in \citeauthor{Hatton+03}. 
Here we are basically 
interested in the density and velocity fields directly traced by dark matter
particles. The $N$-body simulation contains $256^{3}$ particles of mass $8.3
\times 10^9 M_{\odot}$ in a box of 150 Mpc size and it is run with a softening
length amounting to a spatial resolution of 29 kpc.  The simulation was run for
a flat universe with cosmological parameters $\Omega_{0} = 0.333,
\Omega_{\Lambda} = 0.667$, $H_{0}=66.7\,\rm km\,s^{-1}\,Mpc^{-1}$, and
$\sigma_8 = 0.88$.  Once the simulation is run, halos of dark matter are
detected with a `Friends-of-Friends' (FoF) algorithm \citep{DEFW85}, with a
variable linking length such that the minimum mass of the FoF groups is $1.65
\times 10^{11} M_{\odot}$ (20 particles) at any time step. With this method,
over $2 \times 10^{4}$ halos are detected at the final timestep, corresponding
to the present-day ($z=0$) Universe. The {\tt GALICS} halo finder does not
allow halos within halos, so that a cluster, to which is assigned a massive
halo, cannot contain smaller halos within it.

\section{Halo structure in radial phase space}
\label{phase}

%%%%%%%%%%%%%%%%%%%%%%%%%%%%%%%%%%%%%FIG1%%%%%%%%%%%%%%%%%%%%%%%%%%%%%%%%%%%%%%
\begin{figure*}
\centering 
\resizebox{\hsize}{!}{\includegraphics{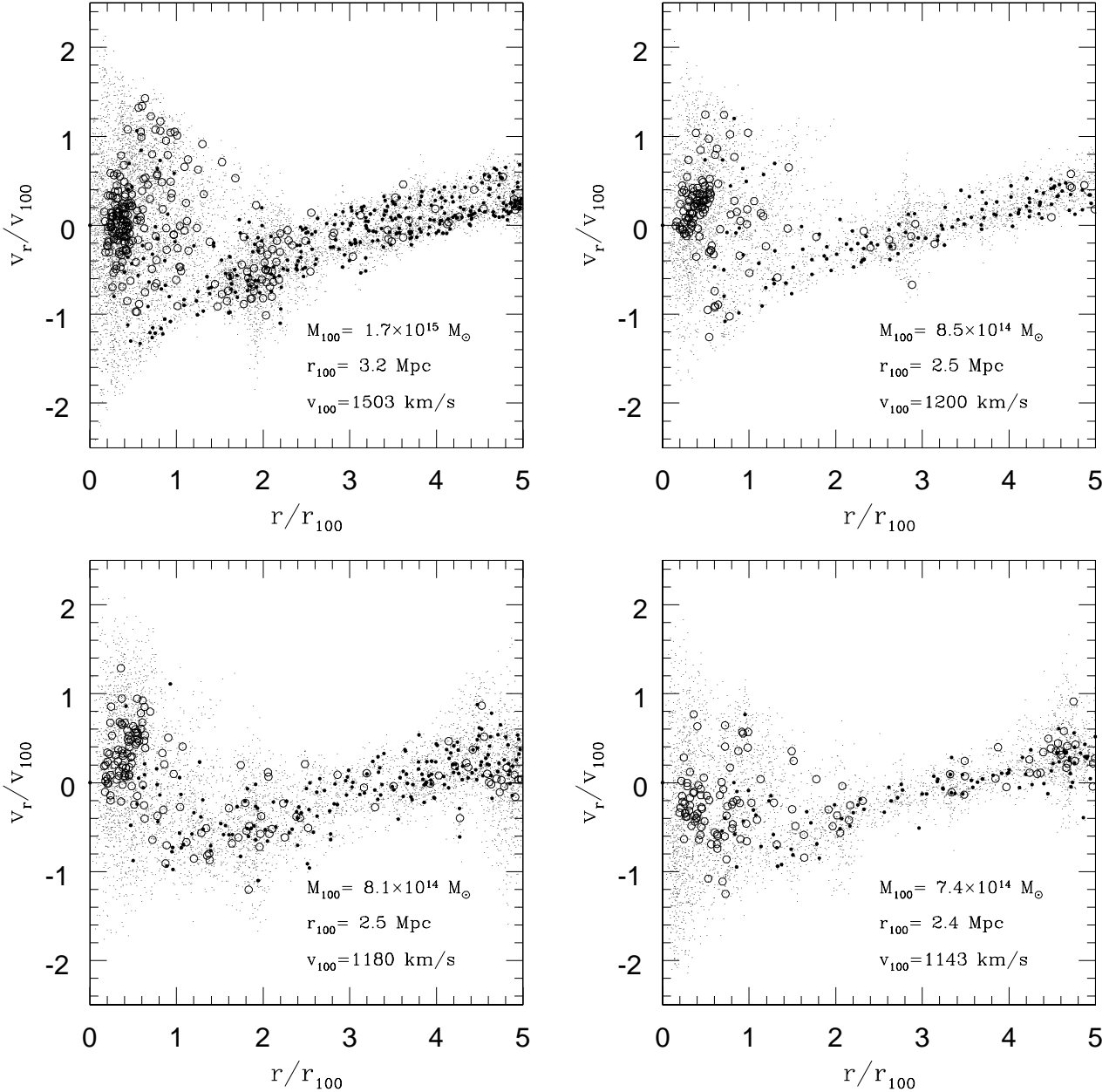}}

\caption{3D radial phase space plots (radial velocity versus radial distance,
both relative to the halo center and normalized to the virial circular velocity
and radius, respectively)  of dark matter particles in a $\Lambda$CDM 
cosmological $N$-body simulation for the four most massive  halos at redshift
$z=0$.   The virial radius,  mass and circular velocity at virial radius are
listed on the right corner of each plot.  The \emph{open} and \emph{small
closed circles} are the identified halos in the cosmological simulation
respectively without and with galaxies within them.
}             
\label{radial}
\end{figure*}
%%%%%%%%%%%%%%%%%%%%%%%%%%%%%%%%%%%%%%%%%%%%%%%%%%%%%%%%%%%%%%%%%%%%%%%%%%%%%%

Fig.~\ref{radial} shows the radial phase space diagrams, i.e. radial velocity
vs. radial distance, both relative to the halo center, for 4  massive halos. We
have studied the final output of the simulation at $z=0$.  The centers of the
isolated halos shown in Fig.~\ref{radial} are provided by the simulation output
and correspond to the barycenter of the FoF groups of particles.  Radial
distances are normalized to the virial radius $r_\mathrm{100}$,  corresponding
to the radius where the mean density is 100 times the critical density of the
Universe. We use $r_{100}$ instead of $r_{200}$ because the former represents
better the virial radius in universes with non zero cosmological constant 
\citep{KS96_ApJ}.  Velocities are normalized to the respective circular
velocities at $r_{100}$.

Most of the massive halos in the simulation have a phase space  diagram 
similar to those for the 4 halos  shown  in Fig.~\ref{radial}, once scaled to
the virial radius and circular velocity at the virial radius. In particular,
all plots show a virialized region for radii smaller than the $r_{100}$
(although this virialization is not perfect, for example see the excess of
positive velocity particles in the lower left plot, presumably caused by a
large group that is bouncing out of the cluster),  and an infalling region with
velocity increasing with radius and asymptotically reaching the (linear) Hubble
flow. One clearly notices groups or small clusters of particles in the outer
regions (e.g. at $r = 3\,r_{100}$ in the upper right plot), which display
\emph{Fingers of God} patterns in phase space. The material within 1 or
$2\,r_{100}$  that bounces out of the cluster should form a pattern symmetric
to the infalling pattern relative to the zero velocity line. It is smeared out
by numerical two-body relaxation (S. Colombi informed us  that the rebounding
region is seen more sharply when simulations are run with increased potential
softening lengths that reduce the numerical relaxation).  The global  aspect of
these phase space plots is similar to those shown by \citeauthor{FM01}
(\citeyear{FM01}, Fig. 21), at various epochs of their cosmological simulations
(run with a standard --- $\Omega_m = 1$, $\Omega_\Lambda = 0$ --- CDM
cosmology).

\section{Maximum rebound radius}
\label{maxreb}

An inspection of Fig.~\ref{radial} shows that dark matter particles  beyond the
virialized core  and outside the infalling/expanding zone of phase space can
reach 2 or 2.5 times the virial radius, but not any further, and moreover come
in groups of particles which appear to be tidally shredded in phase space. In
other words, \emph{particles that cross through the core of a cluster cannot
bounce out beyond 2.5 virial radii}.

Now galaxies are not just particles, but arise within particle condensations
known as dark matter halos, which should arise as vertically-elongated (Fingers
of God) particle condensations in phase space, and therefore ought to avoid the
fairly sparse regions of phase space where the outermost outgoing particles are
seen. The open circles in Fig.~\ref{radial} indicate the halos without galaxies
within them in {\tt GALICS}. The absence of galaxies within halos is a feature
of {\tt GALICS} for halos that cross a larger one (its galaxies become part of
the larger halo). Empty halos can also occur in {\tt GALICS} for isolated halos
in which galaxies have not yet had time to form. The empty halos outside the
infalling/expanding region do not extend beyond $1.7\,r_{100}$ (upper left and
lower right panels, with the former possibly a member of the group at
$1.9\,r_{100}$). In contrast, the normal halos (filled circles in
Fig.~\ref{radial}) outside the infalling/expanding region do not extend as far
from the main halo. The positions of the empty circles in Fig.~\ref{radial}
therefore suggests that  halos crossing the main halo do not bounce out further
than $1.7\,r_{100}$.

This  maximum rebound radius  is consistent with a close inspection of the
right panel of Figure 20 from \cite{FM01}, which shows that  the largest
rebound radii in one of their cosmological simulations, i.e. the largest radii
of a particle that has experienced at least one pericenter, is 2 Mpc, occurring
at the present epoch, for a cluster whose present-day  virial radius
($r_{200}$) is at 1.7 Mpc (see their Table 2). Hence for that particular shell,
the rebound radius is  only 1.2 times $r_{200}$ and an even smaller factor
times $r_{178}$ (the canonical radius for the cosmology used). Given the
$\Omega_m=1$ cosmology used, for which the spherical infall model yields
scale-free growth,  the rebound radius should be proportional to the turnaround
radius, which itself should be proportional to the virial radius, with a time
growth of $r \sim t^{8/9}$ \citep{Gott75}. We checked that the other rebound
radii occurring earlier were even smaller than the scaled expectation of a 2
Mpc radius today.

One can confirm this result through simple analytical arguments. First, if one
identifies the virial radius to the radius where infalling shells meet the
rebounding shells, this will be very close to the rebound radius itself,
defined as the radius where a shell reaches its second apocenter (see Fig.~1 of
\citealp{M92}), so that the rebound radius will be very close to $r_{100}$.

Moreover, one can estimate the rebound radius for a flat cosmology in the
following manner. To begin, assume that the rebound radius is $\widetilde r$
times smaller than the turnaround radius (the first apocenter of the shell),
and occurs at a time equal to $\widetilde t$ times the epoch of turnaround. The
mass within a shell that has reached its second apocenter will be close to but
greater than the mass $M$  within the same shell at turnaround, since some
additional matter will be infalling for the first time. We write this as 
\begin{equation}
\rho_\mathrm{reb} (t_0) \ga \widetilde r^3\,\rho_\mathrm{ta} (T_\mathrm{ta}) \ ,
\label{rhoreb0}
\end{equation}
where the rebound density is for the present epoch, while the turnaround
density is for the epoch of a shell's first apocenter given that its second
apocenter is today.
Given that the mass within the shell at turnaround is
\begin{eqnarray}
M &=& {4\pi\over3} (1+\delta_i)\,\rho_i\,r_i^3 \nonumber \\
&=& {4\pi\over3} \left (1+{\delta_0\over 1+z_i}\right)\,\rho_0
\,r_0^3 \ ,
\end{eqnarray} 
where we used equations~(\ref{rhoofz}) and (\ref{rcomov}),
We then obtain a mean density within the turnaround radius that satisfies
\begin{equation}
{\rho_\mathrm{ta} \over \rho_0} =
\left (1+{\delta_0\over 1+z_i}\right)\,{1\over
y_\mathrm{ta}^3}
\simeq {1\over y_\mathrm{ta}^3} \ .
\label{rhotaoverrho0}
\end{equation}
Equations~(\ref{rhoreb0}) and (\ref{rhotaoverrho0}) 
lead to
\begin{equation}
\rho_\mathrm{reb} \ga \left ({\widetilde r \over y_\mathrm{ta}} \right
)^3\,\Omega_0\,\rho_\mathrm{c,0} \ ,
\label{rhoreb}
\end{equation}
where $\rho_\mathrm{c,0}$ is the present day critical density.
Inverting mean density into mass, assuming $d\ln\rho/d\ln r = -\alpha$, leads
to
\begin{equation}
{r_\mathrm{reb} \over r_\mathrm{vir}} \la
\left ({\Omega_0\over \Delta} \right )^{-1/\alpha}\,\left ({\widetilde r \over
y_\mathrm{ta}} \right )^{-3/\alpha} 
\ ,
\label{radreb}
\end{equation}
where the virial radius $r_\mathrm{vir}$ is defined such that the mean density
within it is $\Delta$ times the critical density.
For \citeauthor*{NFW95} (\citeyear{NFW95}, hereafter NFW)
density profiles, with concentration
parameters $c \simeq 6$, as found by \cite{Bullock+01}
in recent $\Lambda$CDM
cosmological simulations, using $\Delta \simeq 100$, we obtain a slope
$\alpha = 2.4$ at the virial radius.

There are two ways one can reach a large rebound radius today: either with a
large ratio of rebound to turnaround radius for a given shell (i.e. $\widetilde
r$ as small as possible, but it cannot be smaller than unity), or with a
turnaround as late as possible (i.e. $\widetilde t$ as small as possible),
which implies a lower turnaround density, hence a lower rebound density (for
given $\widetilde r$).

In the first case,  the rebound radius could be equal to the initial turnaround
radius, i.e. $\widetilde r = 1$. One then expects that a shell that is
presently at its second apocenter, should have reached turnaround at epoch
$t_0/3$, i.e. $\widetilde t = 3$.  We make use of the reasonably simple
expression for the turnaround radius versus time obtained in the Appendix.
Solving equation~(\ref{h0tta}) for $\delta_0$, 
using equations~(\ref{yta}) and (\ref{phi}) for $y_\mathrm{ta}$ and
$(\Omega_0=0.3,\lambda_0=0.7)$, yields 
$\delta_0 = 2.69$ and $y_\mathrm{ta}=0.229$.
Equation~(\ref{radreb}) and $\alpha = 2.4$ then lead to 
$r_\mathrm{reb} \la 1.8\,r_\mathrm{100}$.

Second, in the spirit of the shell rapidly virializing through violent
relaxation at full collapse, one can assume that the shell reaches its second
apocenter right after its first pericenter (full collapse), i.e. $\widetilde t
\ga 2$, with a radius near half its turnaround radius ($\widetilde r = 2$).
This yields (not surprisingly) $r_\mathrm{reb} = 1.03\,r_{100}$.  The very best
case requires $\widetilde r = 1$ and $\widetilde t = 2$, and leads to
$r_\mathrm{reb} = 2.46\,r_{100}$. Table~\ref{rebtab} summarizes these estimates
of the rebound radius, where $\delta_0$ and $y_\mathrm{ta}$ come from
equations~(\ref{yta}), (\ref{phi}) and (\ref{h0tta}) and
$r_\mathrm{reb}/r_{100}$ from equation~(\ref{radreb}). The last four lines are
for the four particle orbits from the simulations of \cite{FM01} that rebound
within the last 7 Gyr, and for these $\delta_0$ is computed so that the epoch
of the second apocenter is the present time, hence the epoch of turnaround is
$t_0/(t_\mathrm{reb}/t_\mathrm{ta})$.

\begin{table}[ht]
\caption{Rebound radius in different scenarios}
\begin{center}
\begin{tabular}{cccccc}
\hline
\hline
$\widetilde r$ & $\widetilde t$ & $\delta_0$ & $y_\mathrm{ta}$ & $r_\mathrm{reb} /
r_{100}$ \\
\hline
1.0&      3.00&   2.69&  0.229&  1.78 \\
2.0&    2.50&   2.42&  0.258& 0.87 \\
2.0&      2.00&   2.15&  0.296&  1.03 \\
1.0&      2.00&   2.15&  0.296&  2.46 \\
1.5 & 2.50&   2.42&  0.258 & 1.25 \\
2.3 & 2.50 & 2.42 & 0.258 & 0.73 \\
2.4 &  2.17&   2.24&  0.282&  0.77 \\
2.3 &  3.58&   2.99&  0.205&  0.55 \\
\hline
\end{tabular} 
\end{center}
\label{rebtab}
\end{table}

Although, the uncertain effects of relaxation as the shell crosses through the
virialized region lead to uncertain values of the ratios of rebound to
turnaround radius ($\widetilde r$) and time ($\widetilde t$), \emph{in the most
favorable spherical infall model,  the rebound radius cannot be greater than
2.5 times the virial radius}. The cosmological simulations of \citeauthor{FM01}
suggest rebound radii of order of $r_{100}$, but recall that they are for a
different cosmology ($\Omega_0 = 1, \lambda_0 = 0$). The use of real
simulations has the added advantage of incorporating the effects of two-body
encounters that can push material beyond the theoretical rebound radius.
However, it is not always easy to distinguish in a given snapshot of phase
space (e.g. Fig.~\ref{radial}) the material that is bouncing out of a structure
with material that is infalling for the first time (except that the former
particles are in halos without galaxies), and the analysis of particle
histories, done for the simulations of \citeauthor{FM01}, is beyond the scope
of this paper for the {\tt GALICS} simulations.

\section{Virial radius, mass and velocity of the Virgo cluster}
\label{virVirgo}

For the application of results of Sect.~\ref{maxreb} to the Virgo cluster, we require an 
estimate of the virial radius of the cluster. We estimate $r_{100}$ through the 
large-scale X-ray observations of the Virgo cluster
obtained with the {\sf ROSAT All-Sky Survey} by \cite*{SBB99}.
The peak of the X-ray emission in Virgo coincides with the position of the
giant elliptical \object{M87}, and we use the integrated mass profile around
\object{M87} 
obtained by \citeauthor{SBB99} to derive the virial radius.
With their isothermal approximation for the total mass profile (their
Fig.~11a), $M(r)/r$ is independent of radius for $r \gg r_c$ and also of the
assumed distance to Virgo, yielding
\begin{equation}
{G\,M(r)\over r} = 3\,\beta\,{kT\over \mu m_p} \ ,
\label{moverr}
\end{equation}
where $\beta = 0.47$ is the shape parameter of the X-ray gas density profile,
$kT = 2.7\,\rm keV$ and $\mu m_p$ is the mean particle mass, generally
assumed for a hot plasma to be roughly $0.6\,m_p$, where $m_p$ is the proton
mass. 
The virial radius, $r_{100}$, is then obtained through
\begin{equation}
\overline \rho = 
{3 M_{100} \over 4 \pi r_{100}^3} = 
\Delta\,{3\,H_0^2\over 8\,\pi\,G} \ ,
\label{rhovirgo}
\end{equation}
where $\Delta = 100$ is the
spherical overdensity at the virial radius.
Solving equations~(\ref{moverr}) and (\ref{rhovirgo}), one obtains
\begin{eqnarray}
r_{100} &=& {1\over H_0}\,\sqrt{2 \over \Delta}\,\sqrt{3\,\beta\,
kT \over\mu m_p} =
1.65\,h_{2/3}^{-1}\,\rm Mpc \ ,
\label{r100}\\
M_{100} &=& {1\over G\,H_0}\,\sqrt{2\,\over \Delta}\,\left
({3\,\beta\,kT\over \mu m_p}\right )^{3/2} \nonumber \\
&=&
2.30 \times 10^{14}\,h_{2/3}^{-1}\,{\rm M}_{\odot} \ , 
\label{m100}\\
v_{100} &=& \sqrt{G\,M_{100}\over r_{100}} = 
\sqrt{3\,\beta\,kT\over \mu m_p} = 
780 \, \rm km \, s^{-1} \ ,
\label{v100}
\end{eqnarray}
where $v_{100}$ is the circular velocity at $r_{100}$ and 
$h_{2/3} = H_0 / (66.7 \, \rm km \, s^{-1} \,Mpc^{-1})$.

The Virgo cluster is believed to have a complex structure, as it may be
composed of several subclusters \citep*{BTS87} around the elliptical galaxies
\object{M87} (\object{NGC 4486}), \object{M86} (\object{NGC 4406}), and
\object{M49}  (\object{NGC  4472}).  The virial radius of
$1.65\,h_{2/3}^{-1}\,\rm Mpc$ is such that the  important substructure
surrounding \object{M86} is well within it, at a distance of 0.23 (projected)
and 0.44 (3D) times $r_{100}$  (converted to $H_0 = 70 \, \rm km \,
s^{-1}\,Mpc^{-1}$) of \object{M87}.  Similarly, the important substructure
around \object{M49} lies just within the virial radius, at a distance of 0.82
(projected) and 0.88 (3D) times the virial radius of \object{M87}.  Among the
other Messier galaxies in Virgo, which all have fairly secure distances,
\object{M60} is well within the virial radius, \object{M89} lies just at the
virial radius, while \object{M59} is outside.

Analyzing the X-ray emission of the intra-cluster gas, \cite{SBB99} found that
the \object{M49}  subcluster is 2.4 times less massive than the \object{M87}
subcluster. The ratio of masses between the \object{M87} and the \object{M86}
subclusters is even larger  \citep{Bohringer+94}. It therefore appears that the
structure surrounding \object{M87}  is by far the most massive component in the
Virgo cluster, and it is not a bad approximation to choose a single halo to
represent the cluster, so we can apply the results of Sect.~\ref{maxreb}.

\section{Discussion}
\label{disc}

The analysis of Sect.~\ref{maxreb} indicates that the \emph{maximum rebound
radius is between 1 and 2.5 times the virial radius}. Given the virial radius
of 1.65 Mpc for the Virgo cluster, which we derived in Sect.~\ref{virVirgo},
and the maximum rebound radius derived in Sect.~\ref{maxreb}, galaxies passing
through the Virgo cluster core in the past cannot lie further than $(1 - 2.5)
\times 1.65 = 1.7 - 4.1$ Mpc from the cluster center.

\cite{BNM00} found that a very significant fraction of 
particles at a distance between 1 and $2\,r_{200}$ from the centres of
cosmologically simulated clusters have passed through the main body ($r <
r_{200}$) of a cluster progenitor at some earlier epoch.
Note that although the analysis of \citeauthor{BNM00} was performed in terms
of $r_{200}$, they would have very probably gotten similar results had they
scaled their clusters \emph{and progenitors} with $r_{100}$ instead.
As mentioned in Sect.~\ref{intro}, it is not clear from their analysis if
particles can escape beyond $2\,r_{200}$ or even beyond, say, only 
$1.5\,r_{200}$.
Also, it is easier to displace out to large distances particles
rather than
large groups of particles representing a galaxy (or subhalo).
In any event, the result of \citeauthor{BNM00} is consistent with our
analysis of Sect,~\ref{maxreb}.

An examination of Fig.~2 of \cite{Solanes+02} indicates $3\,\sigma$
\HI-deficient galaxies lying between 9 and 30 Mpc from the Local Group, and in
particular galaxies at 10 and 28 Mpc from the Local Group, whose distance error
bars do not reach  the wide range of distances to the Virgo cluster found in
the literature (14 Mpc by \citealp{CJFB98} to 21 Mpc by \citealp{ELTF00}).
Therefore, it appears very difficult to explain  such \HI-deficient galaxies
over 5 Mpc in front or behind the cluster center as having crossed through the
center of the cluster and bounced out if their distance estimates are accurate.
This would suggest that the \HI-deficient galaxies in the outskirts of the
Virgo have not had their interstellar gas ram pressure stripped by the
intracluster diffuse hot gas.

An alternative explanation to the presence on the outskirts of clusters of
\HI-deficient spirals, as well as to the decreased star formation rates and
redder colours 
of galaxies in these regions, relative to field galaxies, is that
the three effects of \HI\ removal, decreased star formation and redder
colours, all intimately linked, may be caused by a significant enhancement of
massive groups of galaxies at the outskirts of clusters, as expected from the
statistics of the primordial density field \citep{Kaiser84} applied to small
groups versus rich clusters \citep{M95_BaltSum}.
If this is the case, we would then expect a correlation between
\HI-deficiency and X-ray emission from the intragroup gas.
However, while tidal effects, which to first order depend on mean density,
regardless of orbit eccentricity \citep{M00_IAP}, 
are similar between less massive groups and more
massive clusters, ram pressure stripping effects, which also depend on the
squared velocity dispersion of the environment, will be much reduced in
groups relative to clusters (e.g. \citealp{AMB99}).

In the companion paper \citep{SMSS03}, we consider different explanations to
the origin of the \HI-deficiency of these outlying galaxies: 1) incorrect
distances, so these objects would in fact lie close enough to the cluster core
to be within the rebound radius and their gas could have been removed by ram
pressure stripping, 2) incorrect estimation of the \HI-deficiencies and 3) 
tidal perturbations (stripping or heating) by nearby companions or within
groups.

\begin{acknowledgements}

We wish to thank St\'ephane Colombi, Yehuda Hoffman, and Ewa {\L}okas  for
useful discussions and Fran\c{c}ois Bouchet, Bruno Guiderdoni and coworkers for
kindly providing us with their $N$-body simulations, and Jeremy Blaizot for
answering our technical questions about the design and access to the
simulations. We also thank an anonymous referee for helpful comments. 
TS acknowledges hospitality of the Institut d'Astrophysique de
Paris where most of this work was done, and she and GAM acknowledge Ewa
{\L}okas for hosting them at the CAMK in Warsaw, where part of this work was
also done. TS was supported by a fellowship of the Ministerio de Eduaci\'on,
Cultura y Deporte of Spain. \end{acknowledgements}

\appendix

\section{Radius, time and density of turnaround in a flat $\Lambda$CDM
cosmology} 
\label{turnaround}

In this appendix, we
compute the parameters of shells at \emph{turnaround}, i.e. reaching their
first apocenter, in a 
$\Lambda$CDM Universe without quintessence ($w_Q=-1$).

In a non-quintessential 
Universe with a cosmological constant, the equation of motion of a shell
of matter is
\begin{equation}
{{\rm d}^2 R\over {\rm d}t^2} = - {G M(R,t) \over R^2} + {\Lambda \over 3}\,R
\ , 
\label{eqmom0}
\end{equation}
where the first term on the right hand side 
is the gravitational force and the second term is the repulsive effect of the
cosmological constant $\Lambda$.
A given shell of matter that first expands with the Universe and then turns
around and collapses will begin to cross shells that have already
settled in a structure only slightly before its own collapse.
Therefore, during the expansion phase, there is no shell crossing and hence
$M(R,t) = \hbox{cst}$.

The equation of motion can then be easily integrated to yield the energy
equation
\begin{equation}
E = 
{1\over 2}\,{\dot r}^2 - {GM\over r} - {1\over 6}\,\Lambda\,r^2
=
{1\over 2}\,{\dot r_i}^2 - {GM\over r_i} - {1\over 6}\,\Lambda\,r_i^2
\,,
\label{energyeq}
\end{equation}
where $E$ is the energy per unit mass of the shell,
and where the $i$ subscript refers to a very early time $t_i$, corresponding to
redshift $z_i \gg 1$ (e.g. $z_i=1000$).
It is convenient to use the quantities
\begin{equation}
\Omega \equiv {8\pi G\rho \over 3\,H^2}
\ ,
\qquad \hbox{and}\qquad
\lambda \equiv {\Lambda \over 3\,H^2} \ ,
\label{omlam}
\end{equation}
which represent the dimensionless mass density and dark energy of the
Universe at any epoch, and we will use subscripts `0' to denote the present
epoch ($z=0$).
In the case of a flat Universe ($\Omega_0 + \lambda_0 = 1$), as confirmed with
the {\sf WMAP} CMB experiment by \cite{Spergel+03}, 
mass and energy conservation applied to the Universe lead to
(e.g. \citealp*{RLT92})
\begin{equation}
H_i^2 = H_0^2\,\left [\Omega_0\,(1+z_i)^3 + \lambda_0\right] 
\label{Hofz}
\end{equation}
At time $t_i$, the mass enclosed within
radius $r_i$ is
\begin{equation}
M = {4\,\pi\over 3}\,\left (1+\delta_i \right) \,\Omega_i\,\rho_{\rm
c,i}\,r_i^3 
= {1\over 2}\,\left (1+\delta_i \right) \,\Omega_i \,H_i^2 \,{r_i^3 \over G}\
,
\label{masseq}
\end{equation}
where $\delta_i$ is the relative overdensity within radius $r_i$ at time $t_i$
(normalized to the density of the Universe at that epoch),
and where we made use of the critical density of the Universe at $t_i$:
\begin{equation}
\rho_\mathrm{c,i} = {3\,H_i^2\over 8\,\pi\,G} \ .
\label{rhoci}
\end{equation}

With a dimensionless radial growth factor $u(r,t) =
r/r_i$,
the first energy equation can be expressed as 
\begin{equation}
\dot u = H_i\,\sqrt{{2\,E \over H_i^2 r_i^2} + {(1+\delta_i)\,\Omega_i\over
u} + \lambda_i\,u^2} \ ,
\label{ueq}
\end{equation}
where we made use of equations~(\ref{energyeq}), 
(\ref{omlam}), 
(\ref{Hofz}), 
and
(\ref{masseq}).
The energy  $E$ of the shell
is obtained by expressing equation~(\ref{ueq}) for the epoch
$t_i$, yielding
\begin{equation}
{2\,E \over H_i^2 r_i^2} = {{\dot {u_i}}^2 \over H_i^2} - \left
[(1+\delta_i)\,\Omega_i + \lambda_i \right ] \ .
\label{energysol}
\end{equation}
The second equality of equation~(\ref{omlam}) yields
\begin{eqnarray}
\lambda_i &=& \left ({H_0\over H_i}\right)^2\,\lambda_0 
= \left ({\Omega_0\over \lambda_0}\,(1+z_i)^3+1 \right)^{-1}
\nonumber \\
&=& O \left((1+z_i)^{-3}\right) \ll 1
\ .
\label{lambdai}
\end{eqnarray}
Mass conservation amounts to a mean density of the Universe that varies as
\begin{equation}
\rho_i \equiv \rho(z_i) = \rho_0\,(1+z_i)^3 \ ,
\label{rhoofz}
\end{equation}
where $\rho_0$ is the present-day mass density of the Universe. This then
yields 
\begin{eqnarray}
\Omega_i &=& \Omega_0\,(1+z_i)^3\,\left ({H_0\over H_i}\right)^2 \nonumber \\
&=& \left [1+{\lambda_0\over\Omega_0}\,(1+z_i)^{-3} \right ]^{-1} 
 = 1 - \lambda_i
\label{omegai}
\end{eqnarray}
Inserting equations~(\ref{energysol}), (\ref{lambdai}) and (\ref{omegai}) into
equation~(\ref{ueq}) yields
\begin{equation}
\dot u = H_i \,\sqrt{{{\dot{u_i}}^2\over H_i^2} + {1+\delta_i-\lambda_i\over
u} + \lambda_i\,u^2 - (1+\delta_i)} \ ,
\label{udot2}
\end{equation}
where we discarded terms in $o(\lambda_i)$.
Now, as pointed out by \cite{Chodorowski88}, \cite*{BES93} and
\cite{Paddy93}, one 
should not assume that the 
initial flow is a pure Hubble flow ($\dot u_i = H_i$), but incorporate the
peculiar motion acquired before $z_i$. 
Using the \cite{Zeldovich70} approximation,
\citeauthor{Chodorowski88} and \citeauthor{BES93} show that  
\begin{equation}
\dot u_i = H_i\,\left (1-{\delta_i \over 3} \right ) \ .
\label{uidot}
\end{equation}
Inserting equation~(\ref{uidot}) into equation~(\ref{udot2}) and discarding
high order terms in $(1+z_i)^{-1}$ and $\lambda_i$ --- writing $\delta_i =
O(1/(1+z_i)) = 
\delta_0/(1+z_i)$ --- yields
\begin{equation}
\dot u = H_i\,\sqrt{-{5\over 3}\,\delta_i + {1\over u} +
\lambda_i\,u^2} 
\ .
\label{udot3}
\end{equation}

The turnaround radius is obtained by solving for $\dot u = 0$, i.e. solving
\begin{equation}
\lambda_i\,u_\mathrm{ta}^3 - {5\over 3}\,\delta_i\,u_\mathrm{ta} + 1 = 0 \ 
\label{cubic0}
\end{equation}
for $u_\mathrm{ta}$.
\cite{LH01} found the same solution without the 5/3 term, as their $\delta_i$
refers to the total initial density, whereas our $\delta_i$ refers to the
growing mode only.
We can go further than \citeauthor{LH01} by expressing and solving a cubic
equation \emph{in terms of present-day quantities}.
With equation~(\ref{lambdai}),
equation~(\ref{cubic0}) reduces to solving
\begin{equation}
{1-\Omega_0\over \Omega_0}\,y_\mathrm{ta}^3 - {5\over
3}\,\delta_0\,y_\mathrm{ta} + 1 = 0
\label{ytaeq}
\end{equation}
for $y_\mathrm{ta} = u_\mathrm{ta}/(1+z_i) = r_\mathrm{ta} / r_0 < 1$, where
\begin{equation}
r_0 = r_i\,(1+z_i)  
\label{rcomov}
\end{equation}
is the comoving radius of the initial perturbation.
The smallest real positive solution of the cubic equation~(\ref{ytaeq}) is
\begin{eqnarray}
y_\mathrm{ta} &=& {2\,\sqrt{5}\over 3}\,
\sqrt{\delta_0\,\Omega_0\over
1-\Omega_0}\,\cos \left ({\phi+\pi\over3} \right )
\label{yta}
\\
\phi &=& \cos^{-1} \sqrt{{729\over 500}\,\left
[{\left(1-\Omega_0\right)/\Omega_0 \over 
\delta_0^3} \right ]}
\label{phi}
\end{eqnarray}
for $\delta_0 \geq 2^{1/3}\,9/10
\,\left[\left(1-\Omega_0\right)/\Omega_0\right]^{1/3}$.

The time of turnaround is then obtained by integrating equation~(\ref{udot3})
and writing $u = (1+z_i)\,y$, yielding
\begin{equation}
{H_i\,T_\mathrm{ta} \over \left (1\!+\!z_i\right)^{3/2}} \!=\!\! 
\int_0^{y_{\rm
ta}}\!\!\!\!\!\!{\sqrt{y}\ {\rm d}y \over
\sqrt{1 \!-\! {(5/ 3)}\,\delta_0\,y \!+\! \left [\left
(1\!-\!\Omega_0\right)/ \Omega_0 \right ]\,y^3}} 
\ .
\label{hitta}
\end{equation}
The starting point of the integration is taken at 0 instead of $1/(1+z_i)$,
which corresponds to adding the negligible time between the $t=0$ and $t_i$.
With equation~(\ref{Hofz}), equation~(\ref{hitta}) becomes
\begin{equation}
H_0\,T_\mathrm{ta} = \int_0^{y_\mathrm{ta}} {\sqrt{y}\ {\rm d}y \over
\sqrt{\Omega_0 - {(5/ 3)}\,\delta_0\,\Omega_0\,y + \left (1\!-\!\Omega_0\right)
\,y^3}} 
\ .
\label{h0tta}
\end{equation}
For $\Omega_0 = 1$, $\lambda_0 = 0$, we get $y_\mathrm{ta} = 3/(5\,\delta_0)$,
$H_0\,T_\mathrm{ta} = (\pi/2)\,[3/(5\delta_0)]^{3/2}$, so that a shell that
collapses today, hence turns around at $H_0\,T_\mathrm{ta} = H_0\,t_0/2  =
1/3$, 
requires a linearly extrapolated density contrast of $\delta_0 =
(3/5)\,(3\,\pi/2)^{2/3} = 1.686$, as expected.

\bibliography{sanchis}

\end{document}